\documentclass[twocolumn,prb,showpacs,floatfix]{revtex4}%
\usepackage[dvips]{graphicx}
\usepackage{epsfig}
\usepackage{psfrag}
\usepackage{amsmath}
\usepackage{amsfonts}
\usepackage{amssymb}
\usepackage{bm}%
\begin{document}
\title{On the theory of resonant susceptibility of dielectric glasses in magnetic
field }
\author{Y. Sereda$^{\dag}$, I. Ya. Polishchuk$^{\ddagger\dag\spadesuit}$,A. L.
Burin$^{\dag\ast}$}
\affiliation{$^{\dag}$Department of Chemistry, Tulane University, New Orleans, LA 70118}
\affiliation{$^{\ddag}$RRC Kurchatov Institute, Kurchatov Sq. 1, 123182 Moscow, Russia}
\affiliation{$^{\spadesuit}$Max-Planck-Institut f\"ur Physik Komplexer Systeme, D-01187
Dresden, Germany}
\date{\today}

\begin{abstract}
The anomalous magnetic field dependence of dielectric properties of insulating
glasses in the temperature interval $10mK<T<50mK$ is considered. In this
temperature range, the dielectric permittivity is defined by the resonant
contribution of tunneling systems. The external magnetic field regulates
nuclear spins of tunneling atoms. This regulation suppresses a nuclear
quadrupole interaction of these spins with lattice and, thus, affects the
dielectric response of tunneling systems. It is demonstrated that in the
absence of an external magnetic field the nuclear quadrupole interaction $b$
results in the correction to the permittivity $\delta\chi\sim\left\vert
b\right\vert /T$ in the temperature range of interest. An application of a
magnetic field results in a sharp increase of this correction approximately by
a factor of two when the Zeeman splitting $m$ approaches the order of
$\left\vert b\right\vert $. Further increase of the magnetic field results in
a relatively smooth decrease in the correction until the Zeeman splitting
approaches the temperature. This smooth dependence results from tunneling
accompanied by a change of the nuclear spin projection. As the magnetic field
surpasses the temperature, the correction vanishes. The results obtained in
this paper are compared with experiment. A new mechanism of the low
temperature nuclear spin-lattice relaxation in glasses is considered.

\end{abstract}

\maketitle

\section{ Introduction}

In 1998, Strehlow et al.\cite{1} discovered and anomalous low-temperature
sensitivity of the dielectric properties of various dielectric glasses to a
magnetic field. This findings continue to attract the attention both of
experimentalists and theorists.\cite{2, 3, 4, 5, 5a, 5b, 5c, 5d, 5e, pnas} As
a result of further investigations, it was found that the effect was
anomalously pronounced for an insulator in the absence of magnetic impurities.

\begin{figure}
[ptb]
\begin{center}
\includegraphics[
height=0.8887in,
width=1.5221in
]%
{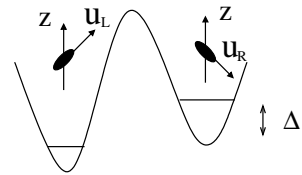}%
\caption{Tunneling system with asymmetry energy $\Delta,$\ and the
corresponding quadrupole quantization axes in the left well $u_{L}$ and in the
right well $u_{R}$\ (see Ref. \cite{9})}%
\label{fig1}%
\end{center}
\end{figure}

Many low-temperature properties of glasses are successfully described by a
standard tunneling model.\cite{6,7,8} In a good approximation the tunneling
systems (TS) can be treated as a particle moving in a double-well potential
(DWP). After observing of the anomalous glassy behavior in a magnetic field,
several extensions of the standard tunneling model have been
suggested.\cite{8a, 8b, 9} In our opinion, the model proposed by
A.~W\"{u}rger, A. Fleischmann, C. Enss\cite{9} is the most viable. It assumes
that the tunneling particle possesses a nuclear electric quadrupole moment. As
a result, the particle energy acquires an extra splitting $b$ in the crystal
electric field gradient (EFG). It is important that, in general case, the
local axis of EFG differ in different wells of DWP (see. Fig. \ref{fig1}). The
magnetic field then interacts with the nuclear spin magnetic moment and
results in the Zeeman splitting. This modifies nuclear spin states in each
well thus affecting the tunneling system properties.

The echo experiments provide convincing evidence of the influence of the
nuclear quadrupole moments on tunneling.\cite{5a,5b,5c,5d} Recently Nagel et
al. \cite{5e} investigated an isotope effect in polarization echo experiments
in glasses. They observed the qualitatively different echo spectra on
amorphous glycerol when hydrogen, which has a \textit{zero} electric
quadrupole moment, was substituted by deuterium, which possesses a
\textit{nonzero} quadrupole moment. Thereupon, one can conclude that the
quadrupole electrical moment of the TS's is the key feature responsible for
the effect.

To our knowledge, there are several theoretical studies of the nuclear
quadrupole effect on glassy dielectric properties.\cite{bodea, PFB,
Burin-hunklinger-Enns, prl2006, prb2006} In papers\cite{prl2006, prb2006} we
have investigated the low temperature limit $T<$ $b.$ We have shown that in
this regime the nuclear quadrupole interaction results in the breakdown of
coherent tunneling due to the formation of a nuclear spin polaron state of
tunneling particle. It was demonstrated that for $T<10mK$ the saturation in
the dielectric constant temperature dependence observed in a variety of
glasses can be explained by that breakdown.

The case of high temperatures $T\gg b$ has been examined in Refs. \cite{bodea,
PFB, Burin-hunklinger-Enns}. Paper\cite{PFB} is concerned with the many - body
relaxation of tunneling system experiencing quadrupole and Zeeman splittings.
Ref. \cite{Burin-hunklinger-Enns} addresses a nonlinear behavior of dielectric
permittivity at high temperatures. Paper \cite{bodea} investigates the
correction to the real part of the permittivity $\delta\chi$ due to the
nuclear quadrupole interaction and the magnetic field. The authors of paper
\cite{bodea} have supposed that this correction is due to tunneling systems,
which have energy splitting of the order of temperature. As a result, a very
weak contribution $\delta\chi$ has been found to possess a strong temperature
dependence $\delta\chi\sim1/T^{6},$ which differs significantly from the
experimental behavior of $\delta\chi\sim1/T$ \cite{bodea,
Burin-hunklinger-Enns}

In the present paper we argue that in the case $\left\vert b\right\vert \ll
T,~$only tunneling systems are important, which have energy splitting of the
order of quadrupole splitting $\Delta,\Delta_{0}\approx b\ll T.$ It is shown
that for vanishing magnetic field the correction to the standard dielectric
permittivity $\chi$ can be expressed as $\delta\chi/\chi\sim\left\vert
b\right\vert /T.$ After application of magnetic field, $\delta\chi$ first
increases sharply. When the Zeeman splitting passes the quadrupole splitting,
$\delta\chi$ reaches the maximum followed by a relatively smooth decrease,
yet, the dependence $\delta\chi\sim\left\vert b\right\vert /T$ remains. The
suggested theory agrees qualitatively with the experimental data.\cite{3, 5c}

The paper is organized as follows. In Section \ref{generalization} we
introduce the tunneling model with the quadrupole and Zeeman interactions. As
an example, we derive the tunneling Hamiltonian for the nuclear spin $I=1.$
Then, in Sec. \ref{permittivity} we investigate the correction to the resonant
permittivity induced by the quadrupole and the Zeeman interactions. In the
following sections we present numerical analysis for the resonant permittivity
in the case of nuclear spin $I=1.$ For the cases $I=1$ and $I=3/2,$ we propose
an exact analytical solution for the permittivity $\delta\chi$ induced only by
the quadrupole interaction. We give the qualitative explanation for the
temperature and the magnetic field dependencies. In conclusion, we consider
the relation of the results obtained to the experimental data. Also, a new
mechanism of the low temperature nuclear spin-lattice relaxation in glasses is
considered, which is due to tunneling system quadrupole interaction.

\section{Generalization of the standard tunneling model\label{generalization}}

According to the standard tunneling model\cite{6,7,8} at low-temperatures,
amorphous solids are represented by an ensemble of tunneling systems. One can
conceive a tunneling system as a particle moving in a double-well potential
(DWP) characterized by the asymmetry energy $\Delta$ and the tunneling
splitting $\Delta_{0}.$ Motion of such a particle is described by the standard
two-level \textit{pseudospin} $1/2$ Hamiltonian
\begin{equation}
h=-\frac{\Delta}{2}\sigma^{z}-\frac{\Delta_{0}}{2}\sigma^{x}. \label{eq:spin}%
\end{equation}
Following Refs.,\cite{6,7,8} we assume that parameters $\Delta,\Delta_{0}$
obey the universal distribution
\begin{equation}
P(\Delta,\Delta_{0})=\frac{P}{\Delta_{0}} \label{distrib}%
\end{equation}
where $P$ is a constant.

The tunneling particle possesses its own internal degrees of freedom
associated with its nuclear spin $\mathbf{I.}$ The tunneling particle gains
Zeeman energy in the magnetic field $\mathbf{B}$
\begin{equation}
E_{int}=-g\beta\mathbf{B\hat{I}},~ \label{emag}%
\end{equation}
where $g$ is the Land\'{e} factor and $\beta$ is the nuclear magneton.
Typically the product $g\beta B$ reaches the value $1mK$ at $B\approx5T.$

Consider the case of the spin $I\geq1$. In this case the nucleus can possess
an electric \textit{quadrupole} moment. It interacts with the crystal field
characterized by the tensor of electric field gradient (EFG) $q_{ij}$. The
Hamiltonian of the particle interacting with the crystal field can be
expressed as\cite{15}
\begin{equation}
H_{Q}=b\left(  I_{1}^{2}+\frac{\varkappa}{3}\left(  I_{2}^{2}-I_{3}%
^{2}\right)  -\mathbf{I}^{2}/3\right)  . \label{f30.03}%
\end{equation}
Here, the parameter $b=\frac{3e^{~2~}Qq_{11}}{4I(2I-1)}$ designates the
quadrupole interaction constant and the asymmetry parameter is given by
\begin{equation}
\varkappa=\frac{q_{22}-q_{33}}{q_{11}}. \label{f30.02}%
\end{equation}
We assume that the Cartesian $e_{1},e_{2},e_{3}$ axes are chosen so that
$q_{33}\leq q_{22}\leq q_{11}$, since then $0\leq$ $\varkappa\leq1.$ If
$\varkappa=0,$ then the EFG possesses axial symmetry. In this case, the
quadrupole energy is completely defined by the spin projection $I_{1}$ and the
quadrupole quantization axis is directed along $e_{1}.$%

For the concreteness we will particularize the case $I=1$ and the case of
axial EFG symmetry $\varkappa=0$. These assumptions simplify the problem
noticeably. At the same time, it provides a good qualitative picture of the
phenomena. In addition, some results will be presented for the case $I=3/2.$

Let us choose the Cartesian reference system $\mathbf{e}_{x},\mathbf{e}%
_{y},\mathbf{e}_{z}.$ Consider the eigenfunctions of the operator
$I_{z},\left\vert -1\right\rangle ,\left\vert 0\right\rangle ,\left\vert
1\right\rangle ,$ as a basis. Then, the Zeeman interaction, which is the same
in both wells, reads%
\begin{equation}
H_{m}=-g\beta B\left(
\begin{array}
[c]{ccc}%
\cos\theta & \frac{\sin\theta}{\sqrt{2}}e^{-i\varphi} & 0\\
\frac{\sin\theta}{\sqrt{2}}e^{i\varphi} & 0 & \frac{\sin\theta}{\sqrt{2}%
}e^{-i\varphi}\\
0 & \frac{\sin\theta}{\sqrt{2}}e^{i\varphi} & -\cos\theta
\end{array}
\right)  _{.} \label{no1}%
\end{equation}
Here the angles $\varphi$ and $\theta$ assign the magnetic field direction.

Let $\mathbf{e}$ and $\mathbf{e}^{\prime}$ be the axis of the $EFG$ in the
left and the right well respectively and $\alpha$ the angle between
$\mathbf{e}$ and $\mathbf{e}^{\prime}.$ Without loss of generality one can
assume that $\mathbf{e}$ and $\mathbf{e}^{\prime}~$belong to the
$\mathbf{x-y}~~$plane and $\mathbf{e}_{x}=\mathbf{e}$. Then, in the right
well
\begin{equation}
H_{R}=H_{Q}\left(  \alpha\right)  =\frac{b}{3}\left(
\begin{array}
[c]{ccc}%
-\frac{1}{2} & 0 & \frac{3}{2}e^{-2i\alpha}\\
0 & 1 & 0\\
\frac{3}{2}e^{2i\alpha} & 0 & -\frac{1}{2}%
\end{array}
\right)  _{,} \label{no2}%
\end{equation}
while in the left well%
\begin{equation}
H_{L}=H_{Q}\left(  0\right)  . \label{no3}%
\end{equation}

Thus, the Hamiltonian of the tunneling particle in the presence of the
quadrupole and Zeeman splittings reads
\begin{equation}
H=\frac{1}{2}\cdot\left(
\begin{array}
[c]{cc}%
H_{L} & \Delta_{0}\cdot\mathbf{I}\\
\Delta_{0}\cdot\mathbf{I} & H_{R}%
\end{array}
\right)  , \label{master}%
\end{equation}
where\textbf{ }$\mathbf{I}$ is a unit matrix of rank $3$ and
\begin{align}
H_{L}  &  =-\Delta\cdot\mathbf{I+}2\mathbf{\cdot}\left(  H_{LQ}+H_{m}\right)
\nonumber\\
H_{R}  &  =\Delta\mathbf{\cdot I+}2\mathbf{\cdot}\left(  H_{RQ}+H_{m}.\right)
\label{master2}%
\end{align}
\newline Parameter $b$ is supposed to be approximately independent on the
position of the tunneling particle within a wells. Also, we assume this
parameter is almost the same in different wells. This is justified by the
experimental data for nuclear quadrupole resonance in amorphous $As_{2}S_{3}$
and $As_{2}Se_{3}$ where only about $10\%$ line widening has been found
\cite{rubinstein}

Almost nothing is known about the angle $\alpha$ between the quadrupole axes
in the DWP except for the case of amorphous glycerol where this angle has been
found to have a specific value \cite{5e}. The analytical approach below (see.
Eq. (\ref{no20})) reveals that there is an insignificant dependence of the
result on this angle, perhaps excluding the cases where $\alpha=0,\pi/2,\pi.$
For this reason, we assume the angle distribution to be uniform on the
interval $0\leq\alpha\leq2\pi.$

In this paper, for the sake of simplicity, we are restricting our
consideration to the single tunneling particle possessing a nuclear spin
$I=1.$ This assumption simplifies the analysis. In general, higher spins and
multiple particles are possibly involved in a single tunneling
system.\cite{PFB, prb2006, Heuer}. We assert, however, that our consideration
captures the important physics. In particular, our assertion is based on the
fact that we have not revealed any qualitative difference between spins $I=1$
and $I=3/2.$ The effect of a multiparticle TLS structure seems to be
insignificant at high temperature where the effect of interest is a small
perturbation and we can treat the nuclear quadrupole interaction for each
particle belonging to a tunneling system independently. Then, the resulting
effect is reduced to a sum of single particle contributions. To take into
account the complex TLS structure one should multiply the quadrupole constant
$b$ by a factor of the number of particles involved in the tunneling system.

\section{Resonant permittivity of a multi - level tunneling system
\label{permittivity}}

In the model considered in this paper the application of the external electric
field makes the asymmetry parameter field dependent
\begin{equation}
\Delta\left(  \mathbf{F}\right)  =\Delta-\mathbf{F}\widehat{{\bm\mu}}
\label{no4}%
\end{equation}
Here $\mathbf{F}$ is the strength of the field, and
\begin{equation}
\widehat{{\bm\mu}}=\frac{{\bm\mu}}{2}\left(
\begin{array}
[c]{cc}%
-\mathbf{I} & 0\\
0 & \mathbf{I}%
\end{array}
\right)  \label{no5}%
\end{equation}
is the dipole-moment operator of the system described by the Hamiltonian
(\ref{master}), so that the dipole moment of the particle in the left well is
$-{\bm\mu/2},$ while in the right well it is ${\bm\mu/2}.$

In our recent paper\cite{prb2006} we have derived a general expression for the
dielectric permittivity for the ensemble of quantum system each having a
discrete energy spectrum $E_{m}$%
\begin{equation}
\chi_{ab}=\sum_{m=1}^{2n}\left(  -\mu_{a}\mu_{b}P_{m}\frac{\partial^{2}E_{m}%
}{\partial\Delta^{2}}+\mu_{a}\frac{\partial P_{m}}{\partial F_{b}}%
\frac{\partial E_{m}}{\partial\Delta}\right)  . \label{eq:chi_1}%
\end{equation}
Here $E_{m}$ are the eigenvalues of the Hamiltonian (\ref{master}); $\mu_{a}$
and $F_{b}-$ are the Cartesian coordinate of the vectors ${\bm\mu}$ and
$\mathbf{F}$ respectively; $n=2I+1,~$e.g.,$~n=3$ for the case of nuclear spin
$I=1;$ $P_{m}$ is the population factor for the state $E_{m}.$

Generally, in the low frequency external electrical field $\mathbf{F\sim}$
$\exp\left(  -2\pi i\nu t\right)  $ the population factor $P_{m}$ can change.
This process, however, is important only when the tunneling system experiences
the relaxation for time $\tau$ smaller then the period of electrical field
oscillations $\nu^{-1}$. In the regime of interest, $T<50mK$ and $\nu
>100Hz,$the time $\tau$ is large enough in comparison with the field
oscillation period $\nu^{-1}$.\cite{DDO2} Accordingly, the relaxation rate of
TLS $\tau^{-1}\sim\nu$ is so small that the population of energy levels cannot
follow the rapidly changing field $\mathbf{F}.$ For this reason, the
population factor remains approximately field independent.\cite{7} Therefore,
the contribution of the second term to permittivity in Eq.(\ref{eq:chi_1})
(usually called a relaxational contribution) vanishes. So the permittivity is
only due to the first term in Eq.(\ref{eq:chi_1}) called a resonant
contribution. Then the population factor $P_{m}$ is given by the
\textit{unperturbed }(in the absence of the weak field $\mathbf{F}$)
equilibrium distribution for the by the field \textbf{ }TLS's%
\begin{equation}
P_{m}=\frac{exp\left(  -\frac{E_{m}}{T}\right)  }{\sum_{\gamma=1}%
^{Z}exp\left(  -\frac{E_{\gamma}}{T}\right)  } \label{no6}%
\end{equation}
The contribution of a single TLS should be summed over all TLS's belonging to
the system. This is equivalent to averaging the susceptibility $\chi_{ab}$ in
Eq. (\ref{eq:chi_1}) over energies, tunneling amplitudes, dipole moments, and
angles between quadrupole axes and the external magnetic field (angles
$\alpha,\theta,\varphi$ in Eqs. (\ref{no1}), (\ref{no2})).

Taking the average over directions and absolute values of TLS dipole moments
is straightforward and we can rewrite the resonant TLS contribution in the
form
\begin{align}
\chi_{a,b}^{res}  &  =\delta_{ab}\frac{\mu^{2}}{3}\chi,~~\text{~}%
\chi=\left\langle \sum_{m=1}^{2n}P_{m}\chi_{m}\right\rangle ,\label{no6.01}\\
\chi_{m}  &  =-\frac{\partial^{2}E_{m}}{\partial\Delta^{2}}. \label{eq:chi_2}%
\end{align}
The parameter $\chi_{m}$ is a function of tunneling parameters $\Delta,$
$\Delta_{0},$ the external magnetic field $\mathbf{B}$, and nuclear quadrupole
interaction $b$ in both wells. The angle bracket in Eq. (\ref{no6.01}) mean
the Gibbs averaging, the averaging over distribution (\ref{distrib}), over
directions of the magnetic field $\varphi$ and $\theta$ and of electrical
field gradients $\alpha$. Thus, the permittivity $\chi$ remains the function
of temperature $T$, the quadrupole interaction $b$ and the magnetic field $B$.
Note that one can ignore the external field $\mathbf{F}$ effect on the local
electric field gradient interacting with the nuclear quadrupole because the
field $\mathbf{F}$ is $5$ to $6$ orders of magnitude smaller then the crystal field.

\section{Permittivity of tunneling systems with quadrupole interaction. The
numerical analysis. \label{low temp}}

Below we analyze the case of high temperature, i.e., the case when%
\begin{equation}
b\ll T. \label{no6.1}%
\end{equation}
It is known that the main contribution to permittivity of tunneling systems in
glasses is due to tunneling systems having $\Delta\sim\Delta_{0}\geq
T$.\cite{7, PFB} On the other hand, it is clearly explained in \cite{PFB} that
merely the magnetic field (in the absence of quadrupole interaction) does not
change the permittivity. Changes can occur only if the quadrupole interaction
exists. Below we argue that, in the absence of the magnetic field, the
correction $\delta\chi$ to permittivity caused by the nuclear quadrupole
interaction is associated with tunneling systems having $\Delta\leq\Delta_{0}$
$\leq b.$

The net quadrupole effect can be described by the difference of two
permittivities $\delta\chi,$ one given by Eq. (\ref{no6.01}) and the other
defined by the same Eq. (\ref{no6.01}) taken at the value $b=0$
\begin{equation}
\delta\chi=\chi-\chi(b=0) \label{no6.11}%
\end{equation}
In fact, it is impossible to \textit{"turn off"} the quadrupole interaction to
experimentally investigate how the quadrupole interaction influences the
permittivity, if any. However, a very strong magnetic filed should inhibit the
quadrupole effect \cite{PFB, prl2006, prb2006} This is due to the fact that
the strong magnetic field creates identical local eigen states, characterized
by a certain nuclear spin projection onto the magnetic field in the wells.
Thus, the influence of the quadrupole effect on the permittivity can be
verified experimentally by measuring the difference
\begin{equation}
\delta\chi=\chi_{\exp}(m=0)-\chi(m=\infty),~~m=g\mu B, \label{no6.2}%
\end{equation}
where $m$ is the Zeeman splitting of nuclear spin energy levels by the
external magnetic field.

The permittivity of the tunneling system is completely defined by the energy
spectrum of the Hamiltonian (\ref{master}). This spectrum strongly depends on
the relation between the nuclear quadrupole interaction $b$ and the Zeeman
splitting $m.$

\begin{figure}
[ptb]
\begin{center}
\includegraphics[
height=1.9389in,
width=2.6609in
]%
{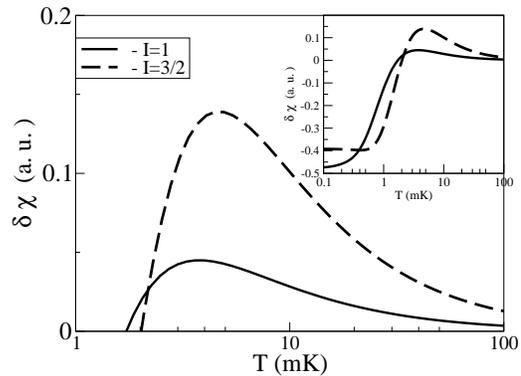}%
\caption{A temperature dependence of the contribution to the permittivity of
tunneling systems due the quadrupole interaction. The graphs clearly shows the
$1/T$ dependence above the temperature $T=5mK.$ The graphs in the inset also
shows a negative contribution below $T=1mK.$}%
\label{fig1(13-06)}%
\end{center}
\end{figure}

\begin{figure}
[ptb]
\begin{center}
\includegraphics[
height=1.672in,
width=2.3416in
]%
{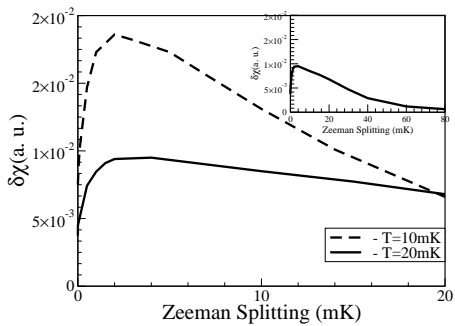}%
\caption{Quasi-plateau magnetic field dependence of the resonant permittivity
for the Zeeman splitting $5mK<m<20mK.$ The graph on the inset displays quasi
exponential $\exp(-b/m)$ dependence for the Zeeman splitting $m>20mK.$}%
\label{fig2(13-06)}%
\end{center}
\end{figure}

Consider first the case $m\ll b$ where one can neglect the magnetic field,
i.e., approximately assuming $m=0.$ Let us first present the results of the
numerical simulation for the permittivity (see. Fig. \ref{fig1(13-06)},
\ref{fig2(13-06)}) in the case of $b=1mK.$These results were obtained as
follows. First, we chosen a certain set of admissible values of the parameters
deterring the Hamiltonian (\ref{master}). Next, we found numerically the eigen
values of the Hamiltonian and used Eqs. (\ref{no6}), (\ref{no6.01}),
(\ref{eq:chi_2}) to calculate the permittivity of a TLS with fixed
parameters.\cite{comment} Then, numerical integration of the single TLS
responses over all relevant parameters was performed to evaluate the effect of
averaging. Fig. \ref{fig1(13-06)} represents the permittivity induced by the
quadrupole interaction. For low temperatures $T<5~mK$ this correction is
\textit{negative}. Also, in this temperature range, as the temperature
increased the permittivity also increased. These "low-temperature results" are
due to the low temperature breakdown of coherent tunneling described in our
previous papers.\cite{prb2006, prl2006}%

Above $5mK$ the quadrupole interaction induced permittivity falls with the
temperature. The analysis reveals that this decreasing follows the dependence
$\left\vert b\right\vert /T.$ Figure \ref{fig1(13-06)} displays this
dependence neatly for both spins $I=1$ and $I=3/2.$

Using a similar numerical approach, one can investigate the magnetic filed
dependence of the permittivity. In this case, the Zeeman splitting has been
taken into account in Eq. (\ref{master}) with averaging the result over the
direction of the magnetic field. Fig. \ref{fig2(13-06)} displays a sharp
increase for the Zeeman splitting region $m<2b,$ then a quasi-plateau in the
magnetic dependence takes place for $2b<m<T$ followed by the decrease for 2
$m>T.$%

Let us now give the physical interpretation of these temperature dependencies.

\subsection{Analytical solution for the permittivity in the case $I=1$}

There are four physically different cases $m\ll b\ll T,$ $m\leq b\ll T,$ $b\ll
m\ll T,$ $b\ll T<m.$

First, let us consider the particular case, when the magnetic field vanishes.
It was assumed and then confirmed by the results obtained in this paper that
only tunneling systems for which $\Delta,\Delta_{0}\leq b$ significantly
contribute to the correction (\ref{no6.2}). For this case, we can assume
$E_{m}\ll T$ in Eq. (\ref{no6}) and, therefore, $e^{-\frac{E_{m}}{T}}%
\approx1-\frac{E_{m}}{T}.$ Since for the Hamiltonian (\ref{master})
\begin{equation}
\sum\limits_{0}^{6}E_{m}=Sp\left(  \hat{H}\right)  =0, \label{no8}%
\end{equation}
one can make the $1/T$ expansion of the correction
\begin{align}
\delta\chi &  \approx\frac{\eta_{1}\left(  \Delta,\Delta_{0},b,\alpha\right)
}{T},\nonumber\\
\eta_{1}\left(  \Delta,\Delta_{0},b,\alpha\right)   &  =\frac{1}%
{6}\left\langle \sum_{m=1}^{6}E_{m}\frac{\partial^{2}E_{m}}{\partial\Delta
^{2}}\right\rangle \label{no9}%
\end{align}
Here the angle bracket mean the same averaging as in Eq. (\ref{no6.01})
excluding Gibbs averaging.

For the Hamiltonian (\ref{master}) one can find exact expressions for $E_{m}$
(see Appendix \ref{appendix1}). Also, the exact analytical averaging procedure
is described in the Appendix. The final expression for the correction to the
permittivity reads%
\begin{equation}
\delta\chi\left(  b,T\right)  =P\frac{|b|}{T}\frac{\pi^{2}}{48}\left(
4-\pi\right)  \approx0.2\cdot P\frac{|b|}{T} \label{no12.01}%
\end{equation}
This dependence shows a nonanalitic $b$- dependence which requires interpretation.%

\begin{figure}
[ptb]
\begin{center}
\includegraphics[
height=1.9993in,
width=3.0007in
]%
{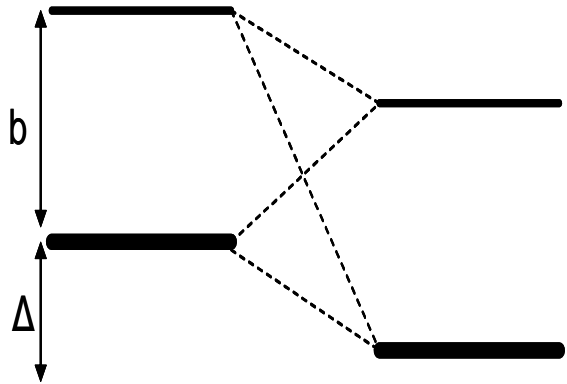}%
\caption{The energy level structure for nuclear spin $I=1$ in the vanishing
magnetic field if the EFG possesses axial simmetry. The thick solid lines
denote the degenerate levels in the left and right wells. The thin solid lines
denote the nondegenerated ones. The dotted lines connect the levels coupled by
the tunneling abplitude $\Delta_{0}.$}%
\label{fig2}%
\end{center}
\end{figure}

Correction to the permittivity $\delta\chi\left(  b,T\right)  $ induced by the
quadrupole interaction $b$ is due to tunneling systems having $\Delta
\approx\Delta_{0}\approx b.$ It is evident that $\delta\chi\left(  b,T\right)
$ can be estimated as $\delta\chi\left(  b=0,T\right)  $ taken for the
tunneling systems having $\Delta\approx\Delta_{0}\approx b.$Thus the
correction $\delta\chi\left(  b,T\right)  $ is estimated to be
\begin{align}
&  \delta\chi\left(  b,T\right) \nonumber\\
&  \approx P\frac{\mu^{2}}{3}\int_{0}^{\left\vert b\right\vert }\frac
{d\Delta_{0}}{\Delta_{0}}\int_{-\left\vert b\right\vert }^{\left\vert
b\right\vert }\frac{\Delta_{0}^{2}d\Delta}{\left(  \Delta^{2}+\Delta_{0}%
^{2}\right)  ^{3/2}}\nonumber\\
&  \times\tanh\left(  \frac{\sqrt{\Delta^{2}+\Delta_{0}^{2}}}{2T}\right)
\nonumber\\
&  \approx\frac{P\mu^{2}\left\vert b\right\vert }{T}. \label{eq:estim1}%
\end{align}
This result correlates with the numerical analysis of Sec. \ref{low temp}.

\subsection{Qualitative estimate of $\delta\chi$ for intermediate,
$m\approx\left\vert b\right\vert ,$ and strong, $m\gg\left\vert b\right\vert
,$ magnetic fields.}

Let us return to the case when tunneling systems are described by the
Hamiltonian (\ref{master}) in the zero magnetic field. If one neglects the
parameter $\Delta_{0},$ the energy spectrum in each of the well consists of
one non-degenerated level and one double-degenerated level (see, Ref.
\cite{PFB} and Fig. \ref{fig2}.

The contribution to the permittivity originates from the left and right states
for which energy detuning is small compared with $\Delta_{0}.$ The tunneling
term mutually bounds either the non-degenerate levels or the degenerate
levels. In this case there exist \textit{three} different values of $\Delta$
that result in resonance. The magnetic field eliminates degeneracy and we
obtain three pairs of mutually coupled levels and there appear \textit{six}
different values of $\Delta$ resulting in resonance. Thus, the effective
number of tunneling systems contributing to the resonance increases (see.
Refs. \cite{PFB, Burin-hunklinger-Enns})%

\begin{figure}
[ptb]
\begin{center}
\includegraphics[
height=1.9993in,
width=3.0007in
]%
{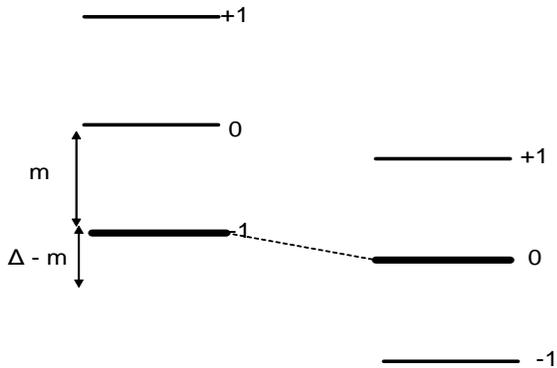}%
\caption{The energy structure of a tunneling system in the case of a strong
magnetic field $b\ll m\,<T.$ The resonance is shown to occur between the state
with nuclear spin projection $I_{z}=-1$ in the left well and the the state
with nuclear spin projection $I_{z}=0$ in the right well. The energy mismatch
is $\left\vert \Delta-m\right\vert \ll m.$}%
\label{fig5}%
\end{center}
\end{figure}

In the case of a half-integer nuclear spin, energy levels in a zero magnetic
field are degenerate according to the Kramers theorem. The application of a
magnetic field eliminates this degeneracy. As in the case of integer nuclear
spin, this results in the increase of the number of resonances. This
circumstance clarifies why the application of the magnetic field increases the
resonant permittivity. Note that at $T<b$ only resonance between the lowest
levels in the left and the right wells is thermodynamically allowed. So the
effect described above vanishes.\cite{prb2006, prl2006}

Let us then turn to the case of $\left\vert b\right\vert \ll m.$ In this case,
one can approximately classify energy levels by nuclear spin projection
$I_{z}.$ If $b=0,$ the coupling between the left and right wells is possible
only for levels having the same nuclear projection. If $b$ is finite, the
overlap factor between the nuclear spin states in the left and the right wells
for levels with different nuclear spin projections is proportional to
$\left\vert b\right\vert /m$ and can be approximated by%
\begin{equation}
\Delta_{0}^{eff}\approx\frac{\left\vert b\right\vert }{m}\Delta_{0}
\label{no18}%
\end{equation}
Resonance between levels with $I_{z}=-1$ in the right well and $I_{z}=0$ in
the left well can be treated as a separate tunneling system having the energy
mismatch $\Delta-m$ and the tunneling amplitude $\Delta_{0}^{eff}\approx
\frac{\left\vert b\right\vert }{m}\Delta_{0}.$ One can estimate its
contribution to the resonant permittivity as \cite{7, 8}
\begin{align}
&  \delta\chi\left(  m,b,T\right) \nonumber\\
&  =P\frac{\mu^{2}}{3}\int_{\left\vert b\right\vert }^{m}\frac{d\Delta_{0}%
}{\Delta_{0}}\int_{-\infty}^{\infty}\frac{\left(  \frac{b}{m}\Delta
_{0}\right)  ^{2}}{\left(  \left(  \Delta-m\right)  +\left(  \frac{b}{m}%
\Delta_{0}\right)  ^{2}\right)  ^{3/2}}\nonumber\\
&  \times\tanh\left(  \frac{\sqrt{\left(  \Delta-m\right)  +\left(  \frac
{b}{T}\Delta_{0}\right)  ^{2}}}{2T}\right) \nonumber\\
&  \approx\frac{P\mu^{2}}{3}\frac{\pi\left\vert b\right\vert }{T} \label{no19}%
\end{align}
This expression results in the following remarkable conclusion. In the case
under consideration, the correction to permittivity is independent of the
magnetic field. This conclusion does not completely correspond to the
numerical simulation (see. Fig. \ref{fig2(13-06)}) where a smooth temperature
decrease in $\delta\chi$ was found. Therefore the result given by Eq.
(\ref{no19}) needs to be amended. While writing Eq. (\ref{no19}) we did not
take into account the population of the level with the nuclear spin projection
$I_{z}=0$ in the right well. This level is separated by an energy gap of the
order of $m$ from the level $I_{z}=-1$ in the right well which is the ground
state of the tunneling system (see Fig. \ref{fig5}). For this reason, the
population of the level $I_{z}=0$ in the right well is approximately
$exp\left(  -m/T\right)  .$ Therefore, the result given by Eq. (\ref{no19})
should be multiplied by the factor $exp\left(  -m/T\right)  $ resulting in a
slow decrease of $\delta\chi$ with increasing $m$ in the region $5mK<m<30mK.$
Comparing Eq. (\ref{no12.01}) and Eq. (\ref{no19}) one concludes that the
correction to the permittivity induced by the magnetic field is several times
larger than for no magnetic field correction. This agrees with the numerical
simulation presented above in Fig. \ref{fig2(13-06)}, where the magnetic field
first causes a sharp increase in the permittivity followed by a slow decrease.
It is clear that the contribution to permittivity given by Eq. (\ref{no19}) is
due to the transition between levels having different spin projections. This
means that the contribution to permittivity is associated with the tunneling
accompanied by a change of the nuclear spin projection. This effect is
discussed below as a new potential mechanism for the spin lattice relaxation.

If the magnetic field is strong enough that $m\gg T,$ only the lowest Zeeman
levels are occupied what brings us and we come to the case of the standard
tunneling model. In this case, the quadrupole interaction effect disappears in
agreement with Fig. \ref{fig5}

\section{CONCLUSION}

It is shown in this paper that at temperatures $T>10mK$ the correction to
permittivity induced by the quadrupole splitting $b$ behaves as $\left\vert
b\right\vert /T.$ This dependence correlates with the experimental data
.\cite{bodea} This comparison with experimental data can be used to estimate
the constant $|b|.$ On the other hand, one can extract from the nuclear
quadrupole resonance (NQR) experiment the frequency of the transition
$\omega_{0}.$ If the tunneling particle carries several atoms, the effect of
the number of the atoms is additive within the perturbation approach. Then,
the ratio $\left\vert b\right\vert /\hbar\omega_{0}$ approximately estimates
the number of atoms per a tunneling system.

Our results exhibit a good qualitative agreement with experiment: there is a
sharp increase in permittivity as the magnetic filed is applied, followed by
slow decrease after the Zeeman splitting passes the value of quadrupole
splitting (see Fig. \ref{fig2(13-06)}). In fact, in $BK7$ glass a similar
effect takes place with position of the maximum at the magnetic field of the
order of Teslas. \cite{3} This agrees with the results obtained in the present
paper. On the other hand, in $BaO-Al_{2}O_{3}-SiO_{2}$ glass the effect is
observed at magnetic field about three orders of magnitude smaller.\cite{2,
5a} The reason for this effect is not clear. This can be due to the presence
of residual paramagnetic impurities interacting with tunneling systems since
the electronic spin in the magnetic field of the order of a few milliTeslas.
acquires the Zeeman energy comparable to the temperature.

However, one can suggest an alternative interpretation of the effect in
$BaO-Al_{2}O_{3}-SiO_{2}.$ In Ref. \cite{PFB} we have shown that resonant
pairs of tunneling systems experiencing quadrupole splitting noticeably
contribute to dielectric loss (imaginary part of the permittivity). It has
been shown that the behavior of these pairs resembles the behavior of
two-level tunneling systems. The effective energy splitting for them is of the
order of $10\mu K.$ This value corresponds to the Zeeman splitting induced by
the magnetic field of the order of milliTeslas. Thus, the magnetic field of
the order of milliTeslas influences the behavior of tunneling pairs affecting
the imaginary part of $\chi$ by means of change in $TLS$ relaxation
rate.\cite{PFB} Because of the Kramers - Kronig relation, this circumstance
should be reflected in the real part of permittivity.

According to the experiment \cite{2, 4, 5a}, the position of the maximum in
the dependence of the dielectric constant $\delta\chi$ on the magnetic field
is sensitive to the value of the external electric field used in the
measurements of the permittivity. This suggests the nonlinear character of the
effect. The interpretation of the nonlinear effect has been proposed in Ref.
\cite{Burin-hunklinger-Enns} It was found that the maximum position is
proportional to the electric field. The dependence on the electric field
qualitatively resembles the behavior in Fig. \ref{fig2(13-06)}. In the present
work, we investigated only the linear response. So, to attain the linear
regime, one should experiment at lower electric fields. The indication to the
linear response is the peak position independence of the electric field(see
Fig \ref{fig2(13-06)}).

It is interesting that the correction to the permittivity at large magnetic
fields $m>b$ is caused by tunneling accompanied by a change in the nuclear
spin projection, Fig. \ref{fig5}. We are justified to anticipate that at low
temperature a similar processes can contribute to spin - lattice relaxation
with the rate exceeding that for dielectric crystals by many orders of
magnitude. This new spin - lattice mechanism should be sensitive to the
resonant transition rate. We expect that this transition is induced by
spectral diffusion \cite{11} which makes the transition an irreversible one.
This relaxation can be slow down by the nuclear spin diffusion needed to bring
a nuclear spin to the TLS neighborhood. \cite{indianpaper} The analysis of
this effect is beyond of the scope of this paper.

\section*{ACKNOWLEDGMENTS}

The work of A. L. Burin, I.Ya Polishchuk, and Yu. Sereda is supported by the
Louisiana Board of Regents (Contract No. LEQSF (2005-08)-RD-A-29) TAMS GL fund
(account no. 211043) through the Tulane University, College of Liberal Arts
and Science. The work of I.Ya Polishchuk also is supported by the program of
Russian Scientific school and Russian Fund for Basic Research. The work of IYP
is supported by the Russian Fund for Basic Researches and the Russian
goal-oriented scientific and technical program "Investigations and
elaborations on priority lines of development of science and technology"
(Contract RI -112/001/526 ).

AB and IYP wish to acknowledge Douglas Osheroff for many useful comments and
for suggesting the possibility of the fast spin-lattice relaxation in glasses
within our model. Also we acknowledge Christian Enss and Siegfried Hunklinger
for many stimulating discussions of experimental data, and Peter Fulde, Alois
W\"{u}rger, Walter Schirmacher and Doru Bodea for the very helpful discussions
of theoretical approaches to the problem. Finally we are all grateful to
organizers and participants of the International Seminar and Workshop "Quantum
Disordered Systems, Glassy Low-Temperature Physics and Physics at the Glass
Transition" (MPIPKS, Dresden, Germany, March 2006) for their attention to our
work and many fruitful remarks.

\section*{APPENDIX}

\label{appendix1}

For the Hamiltonian (\ref{master}) one can find the exact expression for
$E_{m}$. The eigen energies of the Hamiltonian (\ref{master}) in the case
$I=1$ at zero magnetic field read%

\begin{align}
E_{1,2} =b\left(  \frac{1}{3}\pm\frac{E}{2}\right)  ,~E_{3,4,5,6}=b\left(
-\frac{1}{6}\pm\frac{\sqrt{M\pm Y}}{2}\right)  ,\label{no13}\\
E =\sqrt{\left(  \Delta/b\right)  ^{2}+\left(  \Delta_{0}/b\right)  ^{2}},
M=1+E^{2},\nonumber\\
Y=2\sqrt{N}, N=E^{2}-\left(  \Delta_{0}/b\right)  ^{2} \sin^{2}\alpha
.\nonumber
\end{align}
Substituting these expressions into Eq. (\ref{no9}) after simplification one
obtains
\begin{align}
\eta_{1}\left(  \Delta,\Delta_{0},b,\alpha\right) \nonumber\\
\approx\frac{\left(  \Delta/b\right)  ^{2}\left(  \Delta_{0}/b\right)
^{2}\left(  E^{2}+1-4N\right)  \sin^{2}\alpha}{6E^{2}N\left(  4N-(E^{2}%
+1)^{2}\right)  }. \label{no10}%
\end{align}
Note that, if $\alpha=0$ or $\alpha=\pi,$ the quadrupole effect disappears in
agreement with Eq.(\ref{no10}).

Then the total permittivity is given by the expression
\begin{align}
\delta\chi\left(  T,b\right) \nonumber\\
=\frac{P\left\vert b\right\vert }{T}\frac{\mu^{2}}{3}\frac{1}{2}\int_{0}^{\pi
}d\alpha\int_{0}^{T/\left\vert b\right\vert }\frac{d\Delta_{0}}{\Delta_{0}%
}\int_{-T/\left\vert b\right\vert }^{T/\left\vert b\right\vert }%
d\Delta\nonumber\\
\times\eta_{1}\left(  \Delta,\Delta_{0},b,\alpha\right)  . \label{no15}%
\end{align}
To calculate the last integral over $\Delta,\Delta_{0}$ it is convenient to
change the variables as follows.
\begin{equation}
\Delta=E\cdot\sin\varepsilon,~\Delta_{0}=E\cdot\cos\varepsilon;\left\vert
\varepsilon\right\vert \leq\pi/2 \label{no16}%
\end{equation}
Then the correction (\ref{no15}) takes the form
\begin{equation}%
\begin{array}
[c]{c}%
\eta_{1}\left(  E,\varepsilon,b,\alpha\right)  =\frac{c^{2}\cdot\sin
^{2}\varepsilon\cdot\left(  M-4N\right)  }{6g^{2}\cdot\left(  4N-M^{2}\right)
},\\
g=\sqrt{1-c^{2}},~c=\sin\alpha\cdot\cos\varepsilon.
\end{array}
\label{no16.1}%
\end{equation}
The indefinite integral over a parameter $E$ reads \begin{widetext}
\begin{eqnarray}
\int\eta_{1}\left(E,\varepsilon,b,\alpha\right)dE
=\frac{c}{12}\sin
^{2}\varepsilon
\cdot\left(  \arctan\left(  \frac{2\cdot E\cdot c}{1-E^{2}
}\right)  \cdot\left(  1-\left(  \frac{c}{g}\right)^{2}\right)
-\frac{2\cdot c}{g}\cdot\operatorname{arctanh}\left(\frac{2\cdot E\cdot
g}{1+E^{2}}\right)  \right)  , \label{no16.2}%
\nonumber
\end{eqnarray}
\end{widetext}

To calculate the integral (\ref{no15}), one can substitute the integral limit
$T/\left\vert b\right\vert $ by $\infty$ because of the fast integral
convergence. The first term in the rhs of Eq. (\ref{no16.2}) is a broken
function of the argument $E$ at the point $E=1.$ Therefore, integration should
be performed independently over the two intervals $\left(  0,1\right)  $ and
$\left(  1,\infty\right)  .$ After integrating over $E$ one obtains the
intermediate result
\[
\eta_{1}\left(  \varepsilon,\alpha\right)  =\frac{\pi\cdot c}{12}\sin
^{2}\varepsilon\cdot\left(  1-\left(  \frac{c}{g}\right)  ^{2}\right)  .
\]
The contribution of the second term in the rhs of Eq. (\ref{no16.2}) into the
integral (\ref{no15}) vanishes since it turns to zero on the both limits
$E=0,\infty.$

The indefinite integral over $\varepsilon$ reads \begin{widetext}
\begin{eqnarray}
\int\frac{\eta_{1}\left(  \varepsilon,\alpha\right)  }{\cos\varepsilon
}~d\varepsilon
=\frac{\pi}{12}\left(  \cot\alpha
\cdot\left(  \arctan\left(
\frac{\tan\varepsilon}{\cos\alpha}\right)
-2\arctan\left(\frac
{\sin\varepsilon}{1+\cos\varepsilon}\right)  \cdot\cos\alpha\right)
-\sin\varepsilon\cdot\cos\varepsilon\cdot\sin\alpha\right)  .
\nonumber
\end{eqnarray}
\end{widetext}
and
\begin{align}
\eta_{1}\left(  \alpha\right)  =\underset{-\frac{\pi}{2}}{\overset{\frac{\pi
}{2}}{\int}}\frac{\chi_{1}\left(  \varepsilon,\alpha\right)  }{\cos
\varepsilon}~d\varepsilon\nonumber\\
=\frac{\pi^{2}}{12}\cdot\frac{|\cos\alpha|} {\sin\alpha}\cdot\left(
1-|\cos\alpha|\right)  . \label{no20}%
\end{align}
Finally, after the integration over the angle $\alpha$ one obtains

\begin{center}%
\[
\delta\chi\left(  T,b\right)  /\chi=\frac{|b|}{T}\frac{\pi^{2}}{48}\left(
4-\pi\right)  \approx0.2\cdot\frac{|b|}{T}.
\]

\end{center}


\begin{thebibliography}{99}
\bibitem {comment}The numerical intergration over $\Delta$ and $\Delta_{0}$
has been performed for the fixed angles making the exponential substitution
and using the MathLab software to estimate the integral numerically. The
integration over the angles was made using Monte - Carlo approach. The
accuracy of the approach has been checked by comparison with the analytical
result available for the specific case $\theta=\varphi=0.$
\end{thebibliography}
\end{document}